\documentclass{article}

\usepackage{arxiv}

\usepackage[utf8]{inputenc} 
\usepackage[T1]{fontenc}    
\usepackage{hyperref}       
\usepackage{url}            
\usepackage{booktabs}       
\usepackage{amsfonts}       
\usepackage{nicefrac}       
\usepackage{microtype}      
\usepackage{lipsum}
\usepackage{graphicx}
\graphicspath{ {./images/} }

\title{Rethinking Chunk Size for Long-Document Retrieval: A Multi-Dataset Analysis}

\author{
 Sinchana Ramakanth Bhat \\
  Fraunhofer IAIS\\
  Germany \\
  \texttt{sinchana.ramakanth.bhat@iais.fraunhofer.de} \\
   \And
 Max Rudat \\
  Fraunhofer IAIS\\
  Germany \\
  \texttt{max.rudat@iais.fraunhofer.de} \\
  \And
 Jannis Spiekermann \\
  Fraunhofer IAIS\\
  Germany \\
  \texttt{jannis.spiekermann@iais.fraunhofer.de} \\
  \And
 Nicolas Flores-Herr \\
  Fraunhofer IAIS\\
  Germany \\
  \texttt{Nicolas.Flores-Herre@iais.fraunhofer.de} \\
}

\begin{document}
\maketitle
\begin{abstract}
Chunking is a crucial preprocessing step in retrieval-augmented generation (RAG) systems, significantly impacting retrieval effectiveness across diverse datasets. In this study, we systematically evaluate fixed-size chunking strategies and their influence on retrieval performance using multiple embedding models. Our experiments, conducted on both short-form and long-form datasets, reveal that chunk size plays a critical role in retrieval effectiveness - smaller chunks (64-128 tokens) are optimal for datasets with concise, fact-based answers, whereas larger chunks (512-1024 tokens) improve retrieval in datasets requiring broader contextual understanding.
We also analyze the impact of chunking on different embedding models, finding that they exhibit distinct chunking sensitivities. While models like Stella benefit from larger chunks, leveraging global context for long-range retrieval, Snowflake performs better with smaller chunks, excelling at fine-grained, entity-based matching. Our results underscore the trade-offs between chunk size, embedding models, and dataset characteristics, emphasizing the need for improved chunk quality measures, and more comprehensive datasets to advance chunk-based retrieval in long-document Information Retrieval (IR). 
\end{abstract}


\section{Introduction}
\label{sec:introduction}

Retrieval-Augmented Generation (RAG) has emerged as a powerful paradigm in natural language processing (NLP), enabling large language models (LLMs) to enhance response accuracy by incorporating relevant external knowledge retrieved from document corpora \cite{zhang2025jasperstelladistillationsota} \cite{izacard2021leveragingpassageretrievalgenerative}. This approach has significantly improved performance in knowledge-intensive tasks by mitigating the limitations of parametric memory in LLMs and enhancing factual consistency \cite{borgeaud2022improvinglanguagemodelsretrieving}. The effectiveness of RAG systems heavily depend on document chunking strategies, which segment textual data into manageable units before retrieval. Among various chunking techniques, fixed-size token-based chunking remains a prevalent method due to its simplicity and ease of implementation \cite{gao2024retrievalaugmentedgenerationlargelanguage}. Fixed-size chunking segments documents into uniform token-length chunks, ensuring compatibility with transformer-based architectures that have strict token limits \cite{beltagy2020longformerlongdocumenttransformer}.

 However, despite its widespread adoption, the robustness of this approach across varied document lengths remains underexplored. The effectiveness of fixed-size chunking can be influenced by factors such as information dispersion across chunks, redundancy in retrieval, and retrieval precision \cite{liu2023lostmiddlelanguagemodels}. Understanding these factors is crucial to optimizing retrieval performance and downstream generation quality in RAG applications.

 The importance of dataset characteristics in retrieval performance has been well established. Some datasets exhibit high answer locality, where relevant spans are concentrated within a few sentences, while others require reasoning over long contexts spanning multiple chunks \cite{sachan2021end}. Recent studies have highlighted that datasets with diverse document structures demand adaptive chunking strategies to optimize retrieval effectiveness \cite{qi-etal-2022-dureadervis}. Additionally, the role of embedding models in retrieval varies - encoder-based models tend to benefit from exact term matching, whereas decoder-based models leverage broader contextual cues \cite{10.5555/3495724.3496517}.

 Despite its widespread use, fixed-size chunking may not be optimal for all datasets. For instance, hierarchical chunking methods have been shown to improve retrieval by preserving semantic coherence within chunks \cite{liu2021densehierarchicalretrievalopendomain}. Moreover, the challenge of evaluating chunking strategies stems from the difficulty of establishing ground truth relevance in retrieval \cite{qu2024semanticchunkingworthcomputational}. While prior work has explored chunking in traditional retrieval systems, there is limited research on how fixed-size chunking impacts retrieval performance across documents of varying domains in modern RAG architectures. We address this research gap through a structured set of ablations on different domain-specific datasets across different chunk sizes and embedding models. For reproducibility of results, our code is available on Github \footnote{Code and interface description: \url{https://github.com/fraunhofer-iais/chunking-strategies}}. Our key contributions can be summarized as follows: 
\newline
\begin{itemize}
    \item We systematically evaluate the impact of fixed-size chunking on retrieval effectiveness across multiple datasets, analyzing recall@k trends for different chunk sizes and embedding models.
\item We investigate how document length, domain, and dataset diversity influence retrieval performance, providing insights into chunk size selection for different question-answering scenarios.
\item We explore retrieval biases across different embedding models, highlighting how chunking strategies interact with model architectures to optimize retrieval outcomes of chunk size. 
\end{itemize}

\section{Related Work}
\label{sec:related_work} 
\begin{table*}[t]
    \centering
\caption{Overview of dataset statistics, including key properties. Datasets that were stitched together to create longer synthetic documents are marked with an asterisk (*) next to their names.}
    \label{tab:dataset_stats}
    \renewcommand{\arraystretch}{1.1}
    \setlength{\tabcolsep}{4pt} 
    \begin{tabular}{lccccccccc}
        \toprule
        \textbf{Dataset} & \textbf{\# Docs} & \textbf{Q/Doc} & \textbf{Tokens/Doc} & \textbf{Tokens/Q} & \textbf{Tokens/A} &  \textbf{Unique Tokens/Doc}  \\
        \midrule
        NarrativeQA  & 1073  & 1.9  & 51830.4  & 8.5  & 12.6  & 8983.5  \\
        Natural Questions (NQ) & 25010  & 1.0  & 6918.2  & 9.0  & 6.8  & 2864.8 \\
        NewsQA*  & 685  & 11.9  & 8484.5  & 6.5  & 5.2  & 3265.3   \\
        COVID-QA* & 55  & 2.1  & 10009.2  & 8.8  & 11.1  & 2904.5  \\
        TechQA* & 45  & 10.6  & 7597.2  & 51.1  & 46.9  & 2130.1 \\
        SQuAD*  & 306  & 43.7  & 7998.3  & 9.9  & 3.9  & 2949.3  \\
        \bottomrule

    \end{tabular}
\end{table*}

\textbf{\textit{Chunking Strategies}}: The effective segmentation of long documents into manageable chunks is a persistent challenge in information retrieval. We recognize the advancements in dynamic chunking \cite{zhong2024mix} and neural-integrated chunking strategies \cite{sachan2021end}, however, fixed-size chunking remains a widely used and foundational approach due to its simplicity and computational efficiency.  
Furthermore, understanding the behavior and limitations of fixed-size chunking provides a crucial baseline for evaluating the effectiveness of more complex methods. Therefore, we conduct a comprehensive empirical analysis of fixed-size token chunking across a diverse range of modern question answering datasets.
\newline
\textbf{\textit{Embedding Models}}: 
Transformer-based embedding models form the backbone of modern retrieval systems, encoding text into dense representations for similarity-based retrieval. Early works leveraged pre-trained models like BERT \cite{devlin2019bert} and RoBERTa \cite{liu2019roberta} for retrieval, while more recent approaches introduced contrastive learning and hard negative mining to improve retrieval quality \cite{karpukhin2020dense} \cite{izacard2022unsuperviseddenseinformationretrieval}. Dense retrieval models such as DPR \cite{karpukhin2020dense} and Contriever \cite{izacard2021contriever} have demonstrated significant improvements over sparse retrieval methods by leveraging deep contextualized representations. However, embedding models exhibit distinct retrieval biases based on their architectural design. Positional encoding mechanisms play a crucial role in how transformers interpret document structure and chunked inputs. Standard transformers use absolute or sinusoidal positional encodings \cite{vaswani2023attentionneed}, while newer approaches like Rotary Positional Encoding (RoPE) \cite{su2023roformerenhancedtransformerrotary} and Attention with Linear Biases (ALiBi) \cite{press2022trainshorttestlong} modify the handling of positional information. These encodings impact retrieval behavior, especially in chunk-based retrieval settings.
\newline
\textbf{\textit{Dataset Considerations in Chunk-Based Retrieval}}: Understanding the characteristics of question answering datasets is vital for developing robust and generalizable retrieval systems. Recent studies have emphasized the importance of analyzing dataset biases and evaluating model performance across diverse question types \cite{maynez2020faithfulness}. Furthermore, researchers have explored the creation of challenging benchmark datasets that require reasoning over long contexts and multiple documents \cite{tay2020long}. We extend this line of research by conducting a detailed analysis of dataset characteristics and examining their correlation with retrieval performance. Additionally, we investigate the impact of stitching QA pairs within small datasets to form long documents, providing insights into real-world retrieval challenges. This approach is inspired by recent work that has demonstrated the feasibility of stitching datasets for similarity-based evaluations \cite{qu2024semanticchunkingworthcomputational}

\section{Methodology}
\label{sec:methodology}

\textbf{\textit{Chunking and Retrieval}} 
We employ a RAG system built using \textit{LlamaIndex} \cite{Liu_LlamaIndex_2022} to evaluate the impact of chunk size on retrieval performance across various datasets. Our retrieval process begins with fixed-size token splitting, where each document in each dataset is segmented into chunks of predetermined lengths using \textit{LlamaIndex's TokenTextSplitter}. This method ensures consistent chunk sizes, allowing us to isolate the effect of chunk length on retrieval effectiveness. The system first embeds all document chunks using a pre-trained embedding model. Subsequently, a user query is embedded using the same model.
To retrieve the most relevant chunks, cosine similarity is calculated between the query embedding and each chunk embedding. The top k chunks with the highest cosine similarity scores are then selected as the retrieved context.
\newline
\textbf{\textit{Datasets}}
Our study utilizes a diverse set of long extractive question answering datasets, including NarrativeQA \cite{kocisky-etal-2018-narrativeqa}, Natural Questions (NQ) \cite{kwiatkowski-etal-2019-natural}, NewsQA \cite{trischler-etal-2017-newsqa}, COVID-QA \cite{moller-etal-2020-covid}, TechQA \cite{castelli-etal-2020-techqa}, and SQuAD \cite{rajpurkar-etal-2016-squad}. Table \ref{tab:dataset_stats} summarizes the key statistics for each dataset. 
\newline
To ensure the relevance of our evaluation, especially since the challenge is in  finding ground truth data for chunking, we filtered documents by performing a string match comparison between the expected answer and the document text. This step guarantees that the answer is present within the document, focusing our analysis on retrieval accuracy. We also performed cleaning and filtering checks to ensure that each answer appears only once in the document, avoiding ambiguity in our evaluation. 
\newline
A significant challenge in evaluating long RAG datasets is the scarcity of datasets with sufficiently long documents. To address this, we stitched together shorter QA datasets to create longer synthetic documents, ensuring a minimum length of 50,000 characters.

\section{Evaluation}
\label{sec:results}

\begin{table*}[t]
    \centering
    \caption{Retrieval performance (R@K = Recall@K) across different dataset chunk sizes for Stella and Snowflake models.}
    \label{tab:impack of chunk size on ret performance}
    \renewcommand{\arraystretch}{1.1}
    \setlength{\tabcolsep}{4pt} 
    \begin{tabular}{lccccccccccc}
        \toprule
        \textbf{Dataset} & \textbf{Chunk Size} & \multicolumn{5}{c}{\textbf{Stella}} & \multicolumn{5}{c}{\textbf{Snowflake}} \\
        \cmidrule(lr){3-7} \cmidrule(lr){8-12}
        & & \textbf{R@1} & \textbf{R@2} & \textbf{R@3} & \textbf{R@4} & \textbf{R@5} & \textbf{R@1} & \textbf{R@2} & \textbf{R@3} & \textbf{R@4} & \textbf{R@5} \\
        \midrule
        NarrativeQA & 64  & 0.0420  & 0.0685  & 0.0880  & 0.1001  & 0.1079  & 0.0343  & 0.0539  & 0.0712  & 0.0831  & 0.0882  \\
                    & 128 & 0.0571  & 0.0811  & 0.1044  & 0.1260  & 0.1374  & 0.0469  & 0.0701  & 0.0950  & 0.1100  & 0.1204  \\
                    & 256 & 0.0790  & 0.1217  & 0.1510  & 0.1697  & 0.1887  & 0.0616  & 0.0975  & 0.1189  & 0.1469  & 0.1657  \\
                    & 512 & 0.0894  & 0.1360  & 0.1749  & 0.1978  & 0.2263  & 0.0894  & 0.1333  & 0.1657  & 0.1866  & 0.2025  \\
                    & 1024 & \textbf{0.1071} & \textbf{0.1670}  & \textbf{0.2109}  & \textbf{0.2432}  & \textbf{0.2695}  & \textbf{0.1041}  & \textbf{0.1653}  & \textbf{0.2033}  & \textbf{0.2287}  & \textbf{0.2529}  \\
        \midrule
        Natural Questions (NQ) & 64  & 0.1718  & 0.2665  & 0.3338  & 0.3861  & 0.4255  & 0.1500  & 0.2330  & 0.2986  & 0.3464  & 0.3873  \\
        & 128  & 0.2220  & 0.3474  & 0.4295  & 0.4960  & 0.5404  & 0.2040  & 0.3232  & 0.4136  & 0.4808  & 0.5355  \\
        & 256  & 0.3228  & 0.4900  & 0.6043  & 0.6835  & 0.7435  & 0.2634  & 0.4294  & 0.5499  & 0.6333  & 0.6923  \\
        & 512  & \textbf{0.3854}  & 0.5539  & \textbf{0.6689}  & \textbf{0.7448}  & \textbf{0.7954}  & 0.3895  & 0.5817  & 0.6937  & 0.7635  & 0.8125  \\
        & 1024  & 0.3493  & \textbf{0.5561}  & 0.6642  & 0.7273  & 0.7720  & \textbf{0.4774}  & \textbf{0.6835}  & \textbf{0.7937}  & \textbf{0.8581}  & \textbf{0.8982}  \\
        \midrule
        NewsQA* & 64  & 0.3780  & 0.5415  & 0.6328  & 0.6931  & 0.7419  & 0.3204  & 0.4596  & 0.5468  & 0.6083  & 0.6532  \\
                & 128 & 0.4395  & 0.6157  & 0.7179  & 0.7859  & 0.8280  & 0.4156  & 0.5801  & 0.6736  & 0.7334  & 0.7760  \\
        & 256  & 0.4906 & 0.6901 & 0.7978 & 0.8677 & 0.9021 & 0.4448 & 0.6328 &0.7377 & 0.8007 & 0.8370  \\
        & 512  & \textbf{0.5595} & \textbf{0.7869} & \textbf{0.8756} & \textbf{0.9078} & \textbf{0.9276} & 0.5734 & 0.7716 & 0.8398 & 0.8740 & 0.8974  \\
        & 1024  & 0.5202 & 0.6639 & 0.7235 & 0.7663 & 0.7998 & \textbf{0.6668} & \textbf{0.8192} & \textbf{0.8765} & \textbf{0.9110} & \textbf{0.9320}  \\
        \midrule
        COVID-QA* & 64  & \textbf{0.5212}  & \textbf{0.6354}  & \textbf{0.6884} & 0.6975 & 0.7157 & 0.3899 & 0.5278 & 0.5748 & 0.6354 & 0.6672  \\
                  & 128 & 0.4181 & 0.5545  & 0.6551  & 0.6884  & 0.6975  & 0.5324  & 0.6460  & 0.7203  & 0.7475  & 0.7672  \\
        & 256  & 0.4242 & 0.5921 & 0.6648 & 0.7087 & 0.7315 & 0.4087 & 0.5375 & 0.6815 & 0.7709 & \textbf{0.8072}  \\
        & 512  & 0.4060 & 0.5581 & 0.6642 & 0.7390 & 0.7578 & 0.4318 & 0.5739 & 0.6951 & 0.7315 & 0.7487  \\
        & 1024  & 0.3075 & 0.4960 & 0.6684 & \textbf{0.7442} & \textbf{0.7715} & \textbf{0.5421} & \textbf{0.6709} & \textbf{0.7366} & \textbf{0.7745} & 0.8018  \\
        \midrule
        TechQA* & 64  & 0.0486 & 0.0763 & 0.1089 & 0.1413 & 0.1497 & 0.0445 & 0.0771 & 0.0982 & 0.1232 & 0.1418  \\
        & 128  & 0.1650 & 0.2740 & 0.3406 & 0.3680 & 0.3906 & 0.1838 & 0.2848 & 0.3337 & 0.3703 & 0.3885  \\
        & 256  & 0.3995 & 0.5473 & 0.6059 & 0.6717 & 0.6896 & 0.4284 & 0.5826 & 0.6553 & 0.6799 & 0.6838  \\
        & 512  & 0.6138 & \textbf{0.7482} & \textbf{0.8075} & \textbf{0.8677} & \textbf{0.8805} & 0.5811 & 0.7335 & 0.7866 & 0.8213 & 0.8405  \\
        & 1024   & \textbf{0.6192} & 0.7020 & 0.7270 & 0.7834 & 0.8127 & \textbf{0.7154} & \textbf{0.8335} & \textbf{0.8801} & \textbf{0.8980} & \textbf{0.9107}  \\
        \midrule
        SQuAD* & 64  & \textbf{0.6419}  & \textbf{0.7746}  & \textbf{0.8263}  & \textbf{0.8562}  & \textbf{0.8742}  & \textbf{0.6087}  & \textbf{0.7424}  & 0.7955  & 0.8277  & 0.8501  \\
               & 128 & 0.6162  & 0.7542  & 0.8116  & 0.8485  & 0.8712  & 0.6000  & 0.7374  & \textbf{0.8032}  & \textbf{0.8397}  & \textbf{0.8667}  \\
        & 256  & 0.5662 & 0.7103 & 0.7794 & 0.8237 & 0.8506 & 0.5388 & 0.6839 & 0.7583 & 0.8034 & 0.8336  \\
        & 512  & 0.4979 & 0.6518 & 0.7350 & 0.7887 & 0.8263 & 0.4621 & 0.6135 & 0.7023 & 0.7644 & 0.8064  \\
        & 1024  & 0.3855 & 0.5171 & 0.6001 & 0.6596 & 0.7113 & 0.4294 & 0.6044 & 0.7062 & 0.7823 & 0.8368  \\
        \bottomrule
    \end{tabular}
\end{table*}

We evaluate retrieval performance across multiple datasets using a fixed-size chunking strategy. The primary metric is Recall@k, where a retrieval is considered successful if the relevant chunk appears within the top k results. While we report full Recall@k values (k = \{1,2,3,4,5\}), the trends remain conistent across different k, so we primarily focus on Recall@1 in our analysis for brevity. Notable deviations or patterns at higher k are highlighted where relevant.

Each dataset varies in question length, document length, answer locality, and vocabulary diversity, which influences chunking effectiveness (Table \ref{tab:dataset_stats}). To account for these differences, we experiment with multiple fixed chunk sizes: 64, 128, 256, 512, and 1024 tokens using \textit{stella\_en\_1.5B\_v5} \cite{zhang2025jasperstelladistillationsota} and \textit{snowflake-arctic-embed-l-v2.0} \cite{yu2024arcticembed20multilingualretrieval} as embedding models. These experiments were conducted without overlapping tokens.
\subsection{Impact of Chunk Size on Retrieval Performance}
\label{sec:Impact of Chunk Size on Retrieval Performance}
Chunk size plays a significant role in retrieval effectiveness across datasets, impacting how well relevant spans are captured. Our results indicate that smaller chunks (64-128 tokens) perform best for datasets with short, fact-based answers, whereas larger chunks (512-1024 tokens) are necessary for datasets with descriptive or technical responses.
For example, in SQuAD, which has feature concise, entity-based answers, 64-token chunks yield the highest recall@1 which is 64.1\%. However, increasing chunk size reduces recall by 10-15\% at 512 tokens, likely due to excessive context introducing noise. Conversely, NewsQA, with its entity-heavy questions, achieves peak recall@1 at 512 tokens (55.9\%), suggesting that moderate context expansion enhances retrieval without overwhelming the model.
In datasets with long and dispersed answers, such as NarrativeQA, NQ and TechQA, larger chunks significantly improve performance. NarrativeQA recall@1 increases from 4.2\% (64 tokens) to 10.7\% (1024 tokens), highlighting the need for broader context to capture dispersed answer locations. Also, NQ dataset shows similar pattern with highest recalls at 512 and 1024 token lengths respectively. Similarly, TechQA recall@1 improves from 16.5\% (128 tokens) to 61.3\% (512 tokens), demonstrating that wider retrieval windows benefit technical domains where context is essential.
For the COVID-QA dataset, which comprises domain-specific biomedical texts, retrieval performance varies notably with chunk size. While Stella achieves its highest Recall@1 (52.1\%) at 64 tokens, performance gradually declines with larger chunks, suggesting that smaller, focused spans are more effective for this model. In contrast, Snowflake shows a steady improvement, peaking at 1024 tokens with Recall@1 of 54.2\% and Recall@5 of 80.2\%, indicating its stronger ability to leverage extended context without losing relevance.

\subsection{Dataset-Specific Chunking Performance}
\label{sec:Dataset-Specific Chunking Performance}

The impact of chunk size varies significantly across datasets due to differences in document structure, question complexity, answer locality and other factors, which can be inferred from Table \ref{tab:dataset_stats}. Here, we analyze retrieval performance for each dataset in light of these characteristics.

In \textbf{NarrativeQA}, which consists of long, unstructured texts (51830.4 tokens/document), relevant answer spans are often far from the document part that is semantically similar to the question, making small chunks ineffective (recall@1 of 4.2\% at 64 tokens). Performance improves significantly with larger chunks (10.7\% at 1024 tokens), indicating the need for broader context to capture dispersed answer locations. In contrast, \textbf{NewsQA}, with shorter, well-structured news articles (8484.5 tokens/document) and a higher number of questions per document (11.9 on average), performs reasonably well even with smaller chunks across all recalls(37.8\% recall@1 at 64 tokens) and steadily increases with increasing chunk sizes. The structured nature of the text allows relevant spans to be retrieved even with compact representations. In \textbf{NQ}, the chunk size impact is more gradual. Despite relatively short documents (6918.2 tokens/document) and brief answers (6.8 tokens on average), the retrieval performance improves steadily with chunk size, peaking at 1024 tokens for both models. This behavior likely stems from the naturalistic structure of web-sourced documents and the variability in question focus, which often requires incorporating broader context to locate the correct span. Compared to datasets like SQuAD, where answers are tightly localized, NQ exhibits a more distributed answer locality, benefiting from moderate to large chunks without suffering from excessive noise.

For highly domain-specific datasets like TechQA (7597.2 tokens/document) and CovidQA (10009.2  tokens/document), chunking behavior differs due to the nature of the content. \textbf{TechQA} contains long, explanation-heavy answers (46.9 tokens on average) and highly structured text, leading to poor recall at small chunk sizes but substantial improvement as chunk sizes increase (recall@1 rises from 4.8\% at 64 tokens to 71.5\% at 1024 tokens). This suggests that larger chunks are necessary to capture sufficient technical context. In \textbf{COVID-QA}, retrieval performance is shaped by the nature of biomedical literature, moderately long documents (10,009.2 tokens/doc) with relatively verbose answers (11.1 tokens on average) and moderate lexical diversity (2904.5 unique tokens/doc). Compared to datasets like SQuAD, which has very short, factual answers (3.9 tokens) and benefits from compact chunks, COVID-QA spans require more context to capture domain-specific phrasing and nuanced biomedical references. This is reflected in Snowflake's consistent improvement with larger chunks, peaking at 1024 tokens (Recall@1 = 54.2\%). However, Stella achieves its highest Recall@1 (52.1\%) at just 64 tokens, with diminishing returns as chunk size increases, suggesting that it may struggle with noise introduced by extended context. This divergence indicates that while COVID-QA requires moderately sized context windows to retrieve relevant spans, the optimal chunk size is also heavily influenced by model capabilities. Broader context is useful, but only when the retriever can effectively parse dense biomedical language without being overwhelmed by irrelevant content.

In \textbf{SQuAD}, a well-structured Wikipedia-based dataset with a high density of factual questions (43.7 questions per document), retrieval is less sensitive to chunk size. Even at 64 tokens, recall@1 remains high (64.1\%), increasing marginally with larger chunks. This suggests that smaller chunks are sufficient due to the concise nature of the spans that contain the answers (3.9 tokens on average) and the structured format of Wikipedia text.

\subsection{Chunking Impact Across Different Embedding Models}
\label{sec:qualitative-analysis}
Chunking interacts differently with retrieval models, influencing their ability to retrieve relevant spans at varying chunk sizes. As highlighted in  Section \ref{sec:Dataset-Specific Chunking Performance} with COVID-QA, chunk size interacts not only with dataset characteristics but also with the architecture of the underlying retrieval model. Different models exhibit varying sensitivities to chunk size, shaped by their design and pretraining context windows. Stella is a decoder-based model, whereas Snowflake is encoder-based. Hence, they exhibit distinct behaviors due to differences in their training and embedding structures.
Stella demonstrates stronger performance at larger chunk sizes (512-1024 tokens), improving recall@1 by 5-8\% compared to Snowflake in long-document datasets (NarrativeQA, NQ, TechQA). This suggests that Stella benefits from global chunk context, leveraging its large context window of more than 130,000 tokens and corresponding training on similarly long input texts (in the pretraining of its base model Qwen2 \cite{yang2024qwen2technicalreport}). However, at smaller chunk sizes (64-128 tokens), Stella's performance declines, likely due to loss of surrounding context, which weakens retrieval precision.
Conversely, Snowflake maintains competitive performance on small-chunk datasets (SQuAD, CovidQA), where recall@1 remains within 1-2\% of Stella. This highlights its strength in capturing fine-grained entity relationships and handling shorter context windows efficiently. However, its retrieval effectiveness deteriorates at larger chunk sizes, possibly because its pre-training objectives favor local token interactions, likely due to its shorter context window of 8194 tokens.
These findings underscore that chunking effectiveness is model-dependent. While some models excel at long-range retrieval, others are better suited for precise, entity-based matching. Optimizing chunking strategies should take into account model-specific retrieval biases to maximize performance across different datasets.

\section{Limitations}
\label{sec:limitations}
While our study highlights the impact of chunk size on retrieval performance, it has several limitations. Evaluation is based on string matching, which may not fully capture semantic relevance between queries and retrieved chunks. Additionally, while we include a range of datasets, they may not fully reflect real-world information retrieval scenarios - some contain synthetic structures or lack detailed relevance annotations. Finally, our analysis focuses on retrieval performance rather than directly assessing chunk quality, such as coherence or informativeness. Future work should incorporate semantic-aware evaluation metrics, intrinsic chunk assessments, and more realistic benchmarks to address these gaps.

\section{Conclusion}
\label{sec:conclusion}
In this study, we systematically evaluated the impact of fixed-size chunking strategies on retrieval performance across multiple datasets. Our results demonstrate that chunk size significantly influences retrieval effectiveness, with smaller chunks (64-128 tokens) performing well for datasets with concise, fact-based answers, while larger chunks (512-1024 tokens) are necessary for datasets with long, dispersed answers. However, answer complexity alone does not fully explain chunking behavior. Factors such as document structure, question complexity, and answer locality also play critical roles, as reflected in the varying chunk sensitivity across datasets. Furthermore, retrieval performance varies across models, with Stella benefiting from larger chunks due to its ability to capture global context, whereas Snowflake performs better on smaller chunks, leveraging exact phrase retrieval.
Despite these insights, our study highlights several limitations, including reliance on fixed token chunking, string-matching evaluation, and the absence of direct chunk quality measures. Additionally, dataset constraints limit the generalizability of our findings to real-world retrieval tasks. Addressing these challenges through intrinsic chunk coherence
metrics, and more comprehensive datasets will be crucial for advancing chunk-based retrieval models. Ultimately, our findings emphasize the trade-offs between chunk size, retrieval model type, and dataset characteristics. Future work should focus on retrieval-specific embeddings, and fine-tuned evaluation metrics to further optimize retrieval performance in long-document and open-domain question-answering tasks.

\bibliographystyle{unsrt}  
\bibliography{references}  


\end{document}